\documentclass[
superscriptaddress,
amsmath,amssymb,
aps,
preprint,
floatfix
]{revtex4-2}

\usepackage{placeins}

\usepackage[utf8]{inputenc}
\usepackage{graphicx}
\usepackage{dcolumn}
\usepackage{bm}
\usepackage{siunitx}
\DeclareSIUnit\Molar{\textsc{m}}
\DeclareSIUnit\molar{\textsc{m}}

\usepackage{xcolor} 

\definecolor{linkColor}{RGB}{0,70,120}
\usepackage[colorlinks=true, allcolors=linkColor,pdfborder={0 0 0},pdfencoding = auto]{hyperref}

\definecolor{fb_color}{rgb}{0.8,0.5,0.}

\definecolor{sr_color}{rgb}{0.8,0.0,0.6}

\definecolor{ik_color}{rgb}{0.739,0.256,0.08}

\newcommand \ucsbPhys {Department of Physics, University of California at Santa Barbara, Santa Barbara, California 93106, USA}

\newcommand \ucsbBMSE {Biomolecular Science and Engineering, University of California, Santa Barbara, California 93106, USA}

\newcommand \brandeisPhys {The Martin Fisher School of Physics, Brandeis University, Waltham, Massachusetts 02454, USA}

\newcommand \KITP {Kavli Institute of Theoretical Physics, Santa Barbara, California 93106, USA}

\newcommand \JFI {James Frank Institute, University of Chicago, Chicago, IL 60637, USA}

\newcommand \Leinweber {Leinweber Institute for Theoretical Physics, University of Chicago, Chicago, IL 60637, USA}

\newcommand \gtechPhys {School of Physics, Georgia Institute of Technology, Atlanta, GA 30332, USA}

\begin{document}

\title{Active assembly and non-reciprocal dynamics of elastic membranes}

\author{John Berezney}
\thanks{Equal contribution.}
\affiliation{\brandeisPhys}
\affiliation{\ucsbPhys}
 
\author{Sattvic Ray}
\thanks{Equal contribution.}
\affiliation{\ucsbPhys}

\author{Itamar Kolvin}
\thanks{Equal contribution.}
\affiliation{\gtechPhys}

\author{Fridtjof Brauns}
\affiliation{\KITP}
\affiliation{\ucsbPhys}

\author{Sihan Chen}
\affiliation{\Leinweber}
\affiliation{\JFI}

\author{Mark Bowick}
\affiliation{\KITP}
\affiliation{\ucsbPhys}

\author{Seth Fraden}
\affiliation{\brandeisPhys}

\author{Vincenzo Vitelli}
\affiliation{\JFI}
\affiliation{\Leinweber}

\author{Zvonimir Dogic}
\affiliation{\ucsbPhys}
\affiliation{\ucsbBMSE}

\begin{abstract}
\textbf{Equilibrium self-assembly and conventional materials processing techniques fall far short of mimicking dynamic self-actuating processes that are commonplace throughout biology. To bridge the gap between living and synthetic matter, we study adhesive non-thermal fibers immersed in an active fluid. Autonomous chaotic flows power non-equilibrium fiber dynamics, inducing their collisions, generating connections, and weaving a membrane-shaped elastic network. This active assembly generates a hierarchy of shapes, structures, and dynamical processes spanning nanometers to centimeters. Ultimately, it generates an active membrane that exhibits global limit cycles induced by a non-reciprocal coupling between the elastic membrane deformations and the alignment axis of the polar active fluid. Our work merges self-assembly with active matter, demonstrating self-processing materials wherein hierarchical life-like structures and dynamics emerge from an initially structureless suspension.} 
\end{abstract}

\maketitle

\section{Introduction}
During biological development, complex forms and functions emerge from an embryo that lacks large-scale structure. Morphogenesis is underlined by inscribing patterns of in-plane strains into elastic epithelial membranes, causing them to form complex tissues, organs, and whole organisms~\cite{cislo2025morphogenetic,mitchell2022visceral}. Translating such complex yet robust self-organizing processes into the realm of materials science presents an opportunity to generate dynamical materials that are not accessible using paradigms of equilibrium self-assembly and conventional material processing through external energy input. In equilibrium self-assembly, free-energy-minimizing structures emerge from a structureless suspension of microscopic particles with well-defined interactions. Although capable of generating intricate structures~\cite{Nykypanchuk2008, Douglas2009, Jacobs2015, He2020}, self-assembly requires well-tuned conditions and microscopic components that exhibit thermal motion. When local barriers in the free energy landscape exceed the accessible thermal energy, intermediate states become kinetically trapped, and self-assembly does not reach the target structure~\cite{wei2024hierarchical,whitelam2015statistical}. Finally, conventional self-assembly yields only static structures that lack the alluring dynamics and unique mechanical properties observed in living matter. To overcome these limitations intrinsic to equilibrium systems, living matter assembles dynamical structures through a continuous microscopic energy input~\cite{keren2008mechanism,brugues2014physical}.

In synthetic systems, one can generate a non-equilibrium drive with active fluids built from energy-consuming components~\cite{AditiSimha2002, Sanchez2012, wensink2012meso,Zhou2014}. Such materials exhibit autonomous chaotic flows that power the motion of embedded passive objects, allowing them to efficiently explore the phase space~\cite{Wu2000,DiLeonardo2010,Sokolov2010,Sanchez2012}. In turn, such non-equilibrium dynamics can bring together attractive passive particles to generate and grow structures, and once assembled, activate their dynamical soft modes~\cite{Jia2022,Grober2023,pedersen2024active,frechette2025active,palacci2025emergent}. We explore active assembly of passive non-thermal actin fibers powered by microtubule (MT)-based active flows. Adhesive actin filaments and bundles move chaotically in the active flow, colliding with each other, creating permanent connections, and assembling into an elastic membrane. During the ensuing active assembly process, the emergent actin-based elastic structures exert feedback on the MT-based active drive, initiating a multiscale cascade of interacting structural and dynamical processes. Ultimately, we observe the emergence of thin elastic membranes, which exhibit system-size limit-cycle oscillations generated by a non-reciprocal coupling between in-plane actin displacement and MT tilt. These observations reveal the potential of active assembly, where, similar to a biological morphogenesis, complex life-like materials emerge from a structureless mix of active and passive components. 

\begin{figure*}
    \includegraphics[width=0.67\textwidth]{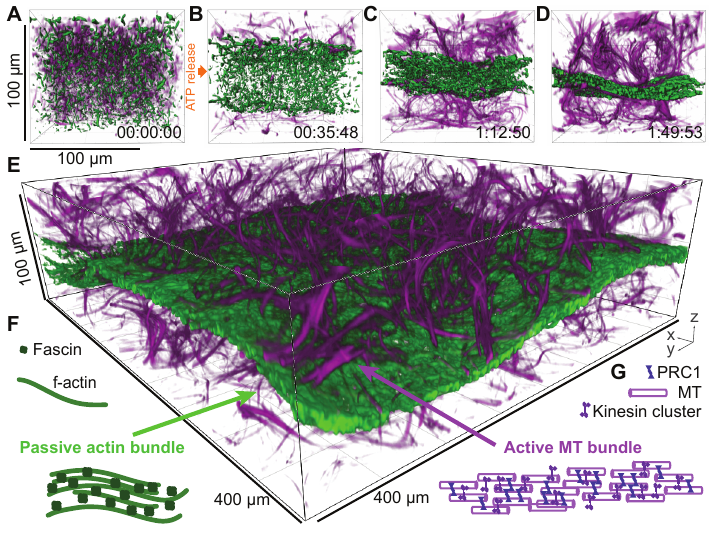}
    \caption{\textbf{Active assembly of elastic membranes} (A-D) Active MT bundles (magenta) drive fascin-linked actin bundles (green) into an elastic membrane.  (E) A larger section of the actin membrane is surrounded by MT bundles. (F) Passive F-actin bundles crosslinked with fascin. (G) The active fluid consists of MTs, MT-specific bundler PRC1, and kinesin-streptavidin clusters. 
    Samples in Active Buffer 1 (SI). Sample in A to E contains \SI{4.0}{\micro\molar} actin.}
    \label{fig:overview}
\end{figure*}

\section*{Active assembly of elastic membranes}
To investigate how active stresses organize adhesive filaments, we constructed a composite system where MT-based active fluid sterically and hydrodynamically interacted with actin and its crosslinker fascin (Fig.~\ref{fig:overview}F, G). Prior to initiation of activity, fascin-crosslinked actin and PRC1-bundled MTs formed a spatially uniform interpenetrating double network that had negligible thermal fluctuations and was static for days~\cite{Berezney2022}. Upon the UV-induced release of caged ATP, kinesin motor clusters (KSA) drove MT bundle extension, leading to persistent chaotic flows. In turn, these flows robustly assembled the actin filaments into a thin elastic membrane at the chamber midplane. Active MT bundles segregated above and below the actin membrane and drove its elastic deformations (Fig.~\ref{fig:overview}A–E, Video~1). The non-equilibrium membrane fluctuations persisted until the ATP was exhausted, demonstrating that the active fluid both assembled an elastic membrane and actuated its dynamics. 

\begin{figure*}
\includegraphics[width=1.0\textwidth]{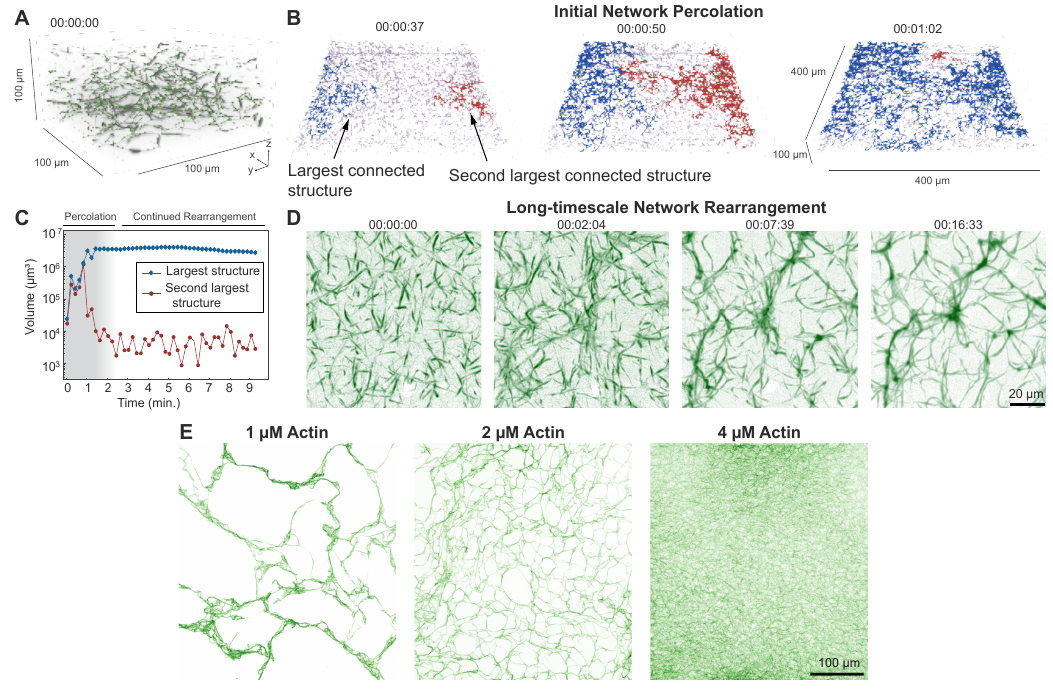}
    \caption{\textbf{Active assembly from preformed bundles.} (A) An initial state consists of well-separated disconnected actin bundles. (B) Activity drives bundle dynamics and rapidly forming connected domains. The largest and second largest connected domains are in blue and red, while others are in gray. (C) Time evolution of the largest and second-largest connected domains. (D) The $z$-projection of the actin network. After geometrical percolation, which occurs within 2 minutes, the network continued coarsening, as evidenced by qualitative changes in its appearance. \SI{1.5}{\micro\molar} actin, \SI{0.5}{\micro\molar} fascin in Active Buffer 1. (E) Snapshots (z-projections) of actin-fascin networks assembled with MT-based active fluid. The fascin to actin ratio is 1:3.} 
    \label{fig:network-formation-fast}
\end{figure*}

We first examined a sample where actin bundles in an initial passive state formed well-separated, locally connected clusters (Fig.~\ref{fig:network-formation-fast}A). Triggering ATP release generated flows that advected actin bundles, causing their enhanced motion, collisions, and formation of irreversible fascin-crosslinked junctions. The intrinsic length scales in this regime revealed the 3D temporal evolution of the network connectivity and cluster size over time. Such analyses revealed that the driven dynamics rapidly increased the size of the largest cluster in the first few minutes, culminating in the formation of a globally percolated structure (Fig.~\ref{fig:network-formation-fast}B-C). Usually, percolation arises from thermally driven cluster aggregation. In comparison, percolation of active bundles is orchestrated by spatiotemporally correlated active flows. After percolation, activity drove further rearrangements, generating a heterogeneous network, in which a fraction of the filamentous material merged, forming brighter and thicker bundles connected through well-defined nodes (Fig.~\ref{fig:network-formation-fast}D and Video~2). 

The network structure generated by active assembly was strongly influenced by the actin concentration (Fig.~\ref{fig:network-formation-fast}E). At low concentrations, the initial passive state consisted of well-separated and highly disconnected actin bundles. Active assembly transformed such configurations into large-mesh networks (Fig.~\ref{fig:network-formation-fast}E left panel, Video~3). Notably, for these conditions, thermal noise alone failed to induce network formation as non-thermal bundles never encountered each other to form connections. We also examined the influence of activity on a network that was fully connected in an initial state. These networks were also rearranged by activity, but less dramatically  (Fig.~\ref{fig:network-formation-fast}E right panel, Video~2). These observations demonstrate how active flows drive geometrical percolation but also determine the steady-state morphologies, delivering final structures that differ dramatically from the initial state. 

\begin{figure}
    \includegraphics[width=\textwidth]{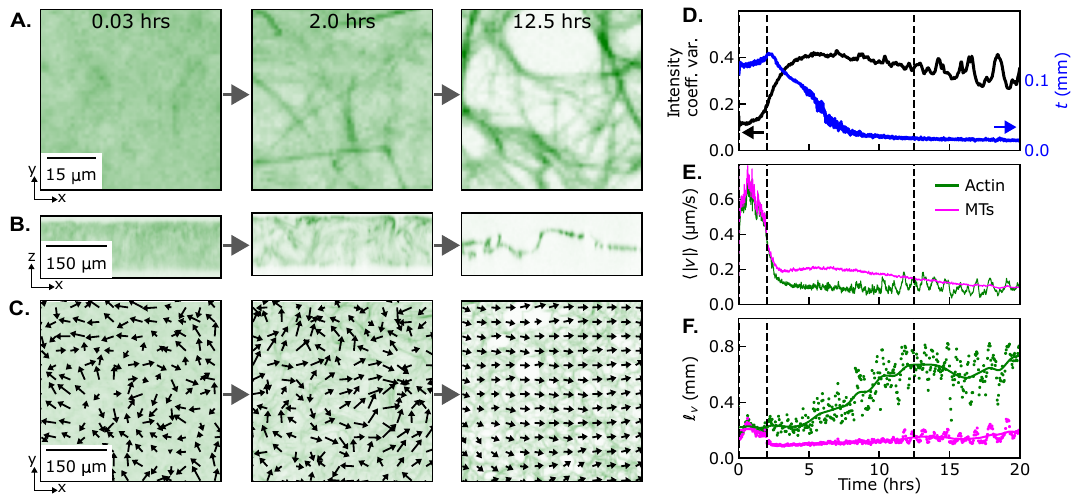} 
    \caption{\textbf{Active assembly from a homogeneous suspension.} (A)~Time evolution of actin intensity $z$-projection normalized by the mean intensity. (B)~Orthogonal slices ($y$-projections) of actin intensity, showing 2D membrane formation. (C) Large-scale membrane flow field overlaid on $z$-projections. Arrows showing velocity are scaled nonlinearly as $\mathbf{v}/|\mathbf{v}|^{0.7}$ for illustration purposes. (D)~Structural changes of actin over time. Intensity coefficient of variation (ratio of standard deviation to mean) of the actin fluorescence is in black. Mean membrane thickness $t$ is in blue. (E)~Time depeendence of spatially averaged mean speed $\langle |\mathbf{v}| \rangle$ of MTs and actin. Between 0--2~h, MTs from the entire volume are included in the average. After 2~h, only external MTs are included. (F)~Velocity correlation length $l_v$ of actin and MT velocity fields. \SI{1.5}{\micro\molar} actin, \SI{1.0}{\micro\molar} fascin in Active Buffer 2.}
    \label{fig:network-formation-slow}
\end{figure}

To explore the applicability of active membrane assembly, we tuned the molecular composition of the passive actin-fascin system to generate an initial state consisting of a homogeneous network of crosslinked filaments, instead of the above-described heterogeneous network of bundles (0.03 hrs - Fig.\ref{fig:network-formation-slow}A, Video~4)~\cite{Lieleg2007,falzone2012assembly}. In this limit, after activation, the actin filaments were first fluidized by chaotic MT flows without forming stable contacts for an extended time. After a lag-time of ~2 hrs, we observed a gradual increase in the variance of the fluorescence intensity across the image, indicative of structure formation (Fig.~\ref{fig:network-formation-slow}D). On comparable time scales, we also observed the appearance of bundle-like structures (2 to 12 hrs- Fig.~\ref{fig:network-formation-slow}A), which became more defined and brighter over time, and the decrease in the membrane thickness (Fig.~\ref{fig:network-formation-slow}B, D). The delayed onset and growth of bundles and the network suggest that bundle formation involves a substantial kinetic barrier that cannot be traversed by thermal fluctuations alone. Active flow enhances the frequency of filament collisions and alignments, enabling the transition into the lower-energy bundled state. 

The network emergence changes the properties of the self-organized active flows. Before bundle formation (0 to 2 hrs), the actin membrane and the surrounding MT-rich regions exhibited fast chaotic flows with comparable speeds (Fig.~\ref{fig:network-formation-slow}C, E). After two hours, concomitant with bundle formation and network contraction, the speeds of both components decreased significantly. In comparison, conventional active fluids maintained steady-state speed throughout their lifetime. This slowdown suggests that active assembly generates passive elastic stresses, which then resist the active flows, resulting in a global slowdown. Intriguingly, over time, the spatial correlations of the active and passive components of the flow fields diverged sharply. The active flows retained short-range correlation lengths of ${\sim}$\SI{100}{\micro\meter} throughout the experiment. The actin membrane initially exhibited a correlation length comparable to that of MTs. However, the membrane length scale progressively increased, reaching ${\sim}$\SI{800}{\micro\meter} after several hours. Thus, short-ranged MT flows drove increasingly longer-ranged membrane motion, providing further evidence for the emergence of the network elasticity. Composed of inextensible actin bundles, the elastic network cannot follow chaotic MT flows, forcing correlated displacements across the membrane. Taken together, these experiments reveal that the active assembly of the membrane is a robust process over a wide range of initial states.

\section{Active buckling}
\begin{figure*}[tb]
    \includegraphics[width=\textwidth]{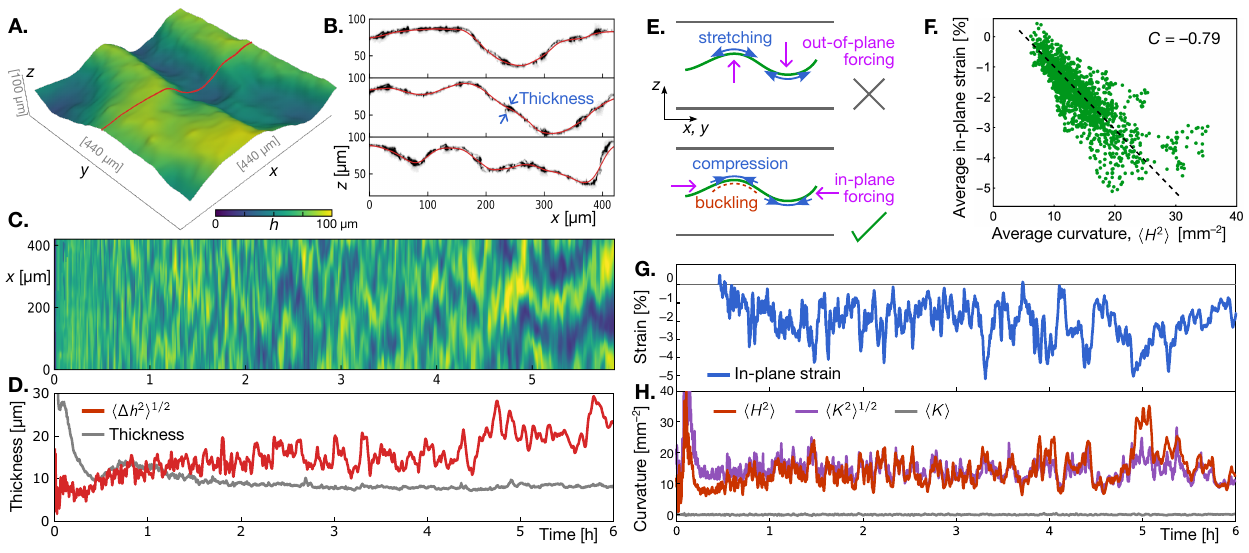}
    \caption{
    \textbf{Active membrane bending and stretching dynamics.}
    (A)~Membrane midsurface reconstructed from a 3D image, \SI{3}{h} after activation. Color indicates the membrane height ($z$-coordinate). 
    (B)~Cross-section, corresponding to the red curve in A, at three consecutive times showing the local actin density (grayscale) and reconstructed midsurface (red lines).
    (C)~Kymograph of the membrane height along the cross-section depicted previously. Height fluctuations exhibit a growing timescale.
    (D)~Time evolution of the average membrane thickness along the local normal, and root-mean-squared height fluctuation.
    (E)~Out-of-plane deformation generated through out-of-plane forcing by MTs (top) or through buckling as a result of in-plane forcing by MTs (bottom).
    (F)~The negative correlation of curvature with strain, suggesting that out-of-plane motion is driven by in-plane forcing. Pearson correlation coefficient $C \approx -0.75$.
    (G)~Time evolution of average in-plane strain. Reference displacement at 27.29~min.
    (H)~Time evolution of the average mean curvature squared $\langle H^2\rangle$, root-mean-squared Gaussian curvature $\langle K^2 \rangle^{1/2}$, and average Gaussian curvature $\langle K \rangle$. All quantities are averaged over the field of view. Same sample as in Figure 2. 
    }
    \label{fig:buckling}
\end{figure*}

Once fully formed, active fluid actuates membrane out-of-plane bending fluctuations (Fig.~\ref{fig:buckling}A,B). To characterize these active fluctuations, we reconstructed the membrane mid-surface $h(x,y,t)$ from the actin fluorescence intensity (Fig.~\ref{fig:buckling}A, B), and quantified its spatially averaged thickness and root-mean-squared height fluctuation(Fig.~\ref{fig:buckling}D). While the thickness rapidly stabilized, the height fluctuations steadily increased in amplitude while their temporal relaxation slowed down~(Fig.~\ref{fig:buckling}C).

The out-of-plane membrane motion is driven by the surrounding extensile MT bundles. Microtubules impinging on the membrane at a steep angle could directly drive out-of-plane motion (Fig.~\ref{fig:buckling}E, top). Alternatively, they could exert in-plane forces, locally compressing the membranes, causing it to bend and buckle (Fig.~\ref{fig:buckling}E, top). To distinguish these two scenarios, we quantified the relationship between the membrane's bending, calculated from the reconstructed mid-surfaces (Fig.~\ref{fig:buckling}H), and its in-plane deformation, calculated from tracked nodes in the actin network (Video~5, Methods). The in-plane strain was dominated by compressions, ranging between $-20\%$ to $+10\%$ but transiently reaching $-40\%$ (Fig.~\ref{fig:strain-analysis}C). The deformations were reversible, suggesting elastic behavior, with the average remaining near zero (Fig.~\ref{fig:buckling}G). Importantly, the in-plane strain was negatively correlated with local bending; more curved membrane sections were more compressed (Fig.~\ref{fig:buckling}F--H), suggesting that bending is due to buckling of the actin membrane under compression. Moreover, the magnitudes of mean curvature and Gaussian curvature nearly matched one another (Fig.~\ref{fig:buckling}H), providing further evidence that the buckling is a result of geometrically incompatible (non-Euclidean) in-plane strains.

\section{Nonreciprocity drives limit-cycle oscillations and nonlinear waves}

\begin{figure*}
    \includegraphics[width=\textwidth]{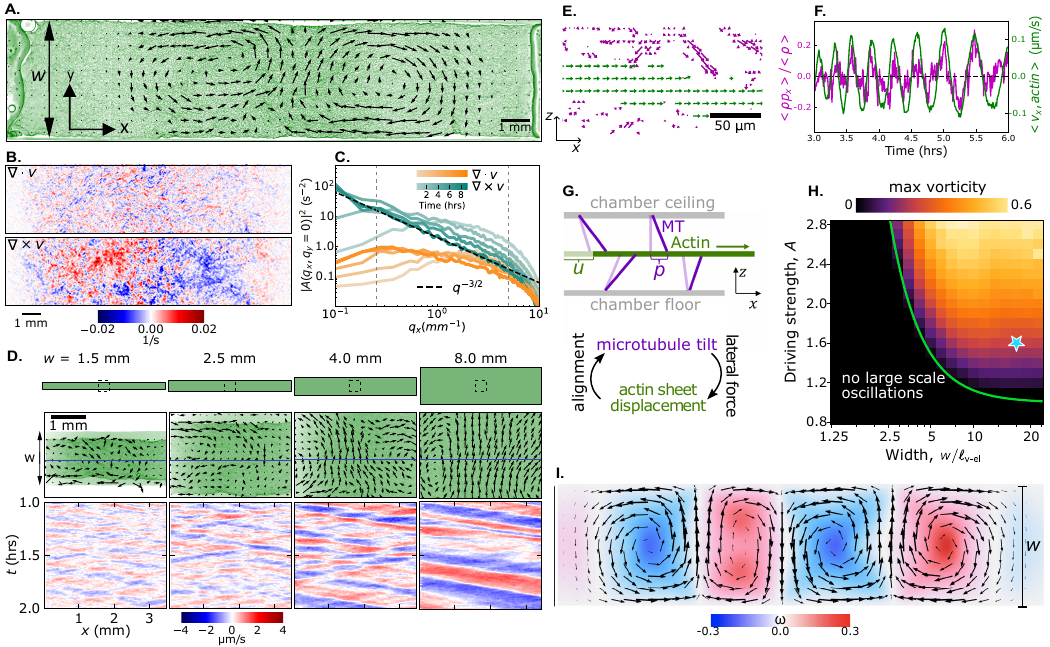}
    \caption{\textbf{System-spanning oscillations.}
    (A)~Centimeter-sized self-excited actin membrane motion. Arrows show the instantaneous velocity field ($t=\SI{6.75}{h}$).  
    (B)~Divergence and vorticity fields for the same time point. 
    (C)~Power spectra of divergence and vorticity fields over time, calculated along the $x$-axis. Each curve is the average spectrum over a 2.2-hour period.
    (D)~Channel width controls self-excited dynamics of the actin membranes. Top: Velocity fields overlaid on actin fluorescence images. Bottom: Kymographs of the $v_y$ component of velocity along the centerline of each sample (blue line). 
    (E)~Polar field of MT bundles calculated from intensity gradients of MTs from $x$-$z$ plane. The green arrow indicates actin membrane motion.  
    (F)~Mean lateral MT polarization $p_x$ (magenta) and mean actin membrane lateral velocity $v_x$ (green) plotted over time.
    (G)~Feedback generates global oscillations: MT bundles align with the membrane motion, which is itself driven by MT forces. The elastic restoring forces provide delayed negative feedback.
    (H)~Linear stability analysis predicts sample-width-dependent oscillations. The green line indicates the threshold for oscillations.
    (I)~Simulation of a nearly incompressible membrane (Video~8).
    (A--C): \SI{3.2}{\micro\molar} actin, \SI{2.2}{\micro\molar} fascin in Active Buffer 2. (D): \SI{3.0}{\micro\molar} actin, \SI{1.0}{\micro\molar} fascin in Active Buffer 1. (E--F): \SI{6.0}{\micro\molar} actin, \SI{2.0}{\micro\molar} fascin in Active Buffer 1.}
    \label{fig:oscillations}
\end{figure*}

The spatial correlation of the in-plane membrane motion increased with time (Fig.~\ref{fig:network-formation-slow}F). Centimeter-sized samples revealed that the long-time outcome of such amplifying dynamics is coherent sample-sized motion, typically consisting of two counter-rotating vortices along the sample's long axis that periodically reversed direction (Fig.~\ref{fig:oscillations}A, Video~6). Such dynamics emerged from temporally uncorrelated fluctuations, as indicated by kymograph of the velocity across the sample (Fig.~\ref{fig:oscillation-kymo}). Accordingly, vorticity became correlated across the entire system (Fig.~\ref{fig:oscillations}B, C). Notably, correlations in the velocity divergence, indicative of stretching and compression, remained short ranged. Thus, at large scales the membrane is effectively incompressible with shear-dominated deformations. Intriguingly, the spectrum of vorticity modes, or enstrophy, follows a $q^{-3/2}$ power law across almost two orders of magnitude (Fig.~\ref{fig:oscillations}C), reminiscent of the inverse energy cascade in 2D turbulence~\cite{paret1997experimental}. However, in contrast to inertial turbulence the dynamics of self-organized membranes is overdamped.

To elucidate the mechanism that powers elastic oscillations, we measured the actin membrane velocity in the $x$ direction and the average tilt of MT bundles relative to the membrane in the $x-z$ plane. The latter yielded a polarization field $p_x$ that encodes the MT tilt direction (Fig.~\ref{fig:oscillations}E). The spatially-averaged MT tilt $\langle p_x \rangle$ and actin membrane velocity $\langle v_x \rangle$ had in-phase sinusoidal oscillations (Fig.~\ref{fig:oscillations}F), suggesting that the interplay between these variables drives the oscillations. Tilted extensile MTs drive lateral membrane motion, which, in turn, further increases the MT tilt, causing a positive feedback (Fig.~\ref{fig:oscillations}G). With increasing displacement, elastic stresses build up, eventually becoming stronger than the active driving force. The membrane then reverses its motion and subsequently entrains the MT polarity anew until elastic stresses again overcome the driving force, leading to another reversal, thus completing the oscillation cycle.

The MT-membrane interactions are captured by a minimal model that non-reciprocally couples the in-plane displacement $u$ of a membrane patch to the local polarization $p$ describing the MTs direction. The balance of forces acting on the overdamped membrane patch reads $\gamma \partial_t u = -k_\mathrm{eff} u + f_0 p/(1+|p|/p_0)$, which captures friction due to drag with the surrounding fluid, a spring-like restoring force accounting for the stresses that build up in the membrane, and a driving force exerted by the extensile MT bundles. The non-linearity in the denominator of the driving term saturates the force at large $|p|$. The MT tilt direction $p$ aligns with the membrane motion and decays due to chaotic dynamics: $\partial_t p = -p/\tau_\mathrm{p} + \alpha \partial_t u$ with the alignment coefficient $\alpha$ and decay timescale $\tau_\mathrm{p}$. Rewriting these equations in matrix form yields
\begin{equation} \label{eq:minimal-model}
    \gamma \partial_t \! \begin{pmatrix} u \\ p \end{pmatrix} = 
    \begin{pmatrix}
        -k_\mathrm{eff} & \frac{f_0}{1+|p|/p_0} \\
        -\alpha k_\mathrm{eff} & \frac{\alpha f_0}{1+|p|/p_0} - \gamma/\tau_\mathrm{p}
    \end{pmatrix} 
    \begin{pmatrix} u \\ p \end{pmatrix}
\end{equation}
In the linear regime ($|p|\ll p_0$), the off-diagonal components with opposite signs indicate anti-symmetric coupling between $u$ and $p$, corresponding to ``run-and-chase'' interactions after rescaling of the variables. When $\alpha f_0 > k_\mathrm{eff} + \gamma/\tau_\mathrm{p}$, the system is linearly unstable and perturbations grow. The non-linear saturation of the active force eventually results in the limit-cycle oscillations of the local displacement and MT polarization observed in experiments (Fig.~\ref{fig:limit-cycles}). This mechanism is analogous to self-aligning active matter systems, where a polar driving field aligning with local velocity causes collective motion \cite{shimoyama1996collective,Baconnier2022,Baconnier2025} and more broadly the emergence of dynamical phases in non-reciprocal many-body 
systems~\cite{you2020nonreciprocity,scheibner2020odd,saha2020scalar,fruchart2021non,tan2022odd,gu2025emergence}. 

The minimal model explains the local oscillatory motion, but doesn't describe spatial properties. We extended the model to 2D (SI). In short, the restoring force $-k_\mathrm{eff} u$ is replaced by an elastic stress $\sigma_{ij}^\mathrm{el} = 2 \mu u_{ij} + (B - \mu) u_{kk} \delta_{ij}$, where $u_{ij} = (\partial_i u_j + \partial_j u_i)/2$ is the strain tensor and $\mu$ and $B$ are respectively the shear and bulk moduli. The motion of the membrane relative to the chamber floor and ceiling induces shear in the surrounding fluid, leading to an effective drag force and viscous stress $\sigma_{ij}^\mathrm{visc} = 2 \eta \partial_t u_{ij} + (\eta_\mathrm{B} - \eta) \partial_t u_{kk} \delta_{ij}$, where shear and bulk viscosities $\eta$, $\eta_\mathrm{B}$ are derived using a lubrication approximation (SI). The incompressibility of the bulk fluid generates an effective membrane drag coefficient that is four times higher for compressional modes than for vortical modes. This effective friction arises because compressive deformation of the membrane drags along the surrounding fluid, which then recirculates from the point where the compressive membrane motion converges (Fig.~\ref{fig:effective-drag}). The recirculating flow generates a high shear rate along the $z$ direction. Thus, compressive deformations are suppressed on large length scales, as observed in experiments (Fig.~\ref{fig:oscillations}A--C).

The spatially extended model reproduces the experimentally observed oscillations with wave-like character that propagate along the channel (Fig.~\ref{fig:oscillations}I, Video~7).  A dispersion relation obtained from the linear stability analysis indicates the growth of wave-like modes (Fig.~\ref{fig:disp-rel}). Importantly, these waves are distinct from sound waves originating from inertia and elasticity; they are driven by activity in an overdamped system, with non-linearities determining the wave amplitude. Furthermore, linear stability analysis reveals that elastic stresses suppress shorter-wavelength modes, so that the longest-wavelength mode grows the fastest (Fig.~\ref{fig:disp-rel}). This explains the transfer of energy from the small-scale driving by MT bundles to the system-size oscillations that dominate the membrane motion. 

Only waves larger than a critical wavelength $\ell_\mathrm{c} = 2\pi
\ell_\text{v-el} / \sqrt{A - 1}$ are unstable. Here, $A = \alpha f_0 \tau_\mathrm{p}$ is the dimensionless driving strength and $\ell_\text{v-el} = \sqrt{(\mu \tau_\mathrm{p} + \eta)/\gamma}$ screening length of viscoelastic stresses in the membrane. Consequently, large-scale unstable modes emerge only if the channel width $w$ is larger than half the critical wavelength $\ell_\mathrm{c}/2$. Since longitudinal modes are suppressed by incompressibility, the channel length is irrelevant for the onset of instability. Thus, the emergence of large-scale waves requires sufficiently wide channels (Fig.~\ref{fig:oscillations}H). To test this prediction, we studied membranes in channels of various widths (Video~8). The presence of wave-like membrane motion was captured by space-time kymographs of $v_y(x,t)$. Robust traveling wave dynamics were only observed for samples with a width above \SI{4}{mm}, corroborating the prediction of the spatially extended model (Fig.~\ref{fig:oscillations}D). 

\section{Discussion}
Our results demonstrate the potential of active assembly, in which life-like materials with complex structure, mechanics, and dynamics emerge from a structureless suspension of active and passive components. Previous studies examined how molecular motors contract filamentous networks, but these lacked steady-state dynamics~\cite {kohler2011structure,Murrell2012,Alvarado2013, foster2015active,Lee2021,Livne2024}. Separating the active drive from the passive stresses greatly expands the range of accessible structures, while also generating long-term non-equilibrium dynamics. Our results connect several scientific themes. When combined with previous studies~\cite{Hemingway2015,Liu2021,Baconnier2022}, our work suggest that self-excited waves and oscillations are ubiquitous features of active solids. The active fluid generates elastic membranes and actuates their non-thermal out-of-plane fluctuations. As exemplified by graphene, such fluctuations greatly increase the effective membrane bending rigidity while decreasing their in-plane stretching modulus~\cite{nelson1987fluctuations,Blees2015}. Analogously, non-thermal fluctuations should also control the mechanics of driven membranes, raising an intriguing scenario where the increasing out-of-plane fluctuations soften the effective in-plane stiffness, thus making the membrane more susceptible to global oscillations. This requires more details models that couple out-of-plane and in-plane deformations. In a complementary direction, there are intriguing connections with previous work on turbulence. further theory work is needed to quantitatively understand the energy cascade underlying the power-law vorticity spectrum and elucidate similarities and differences with classical 2D turbulence~\cite{paret1997experimental}. Our work also demonstrates an experimental route, inspired by morphogenesis, for building active membranes made of elastic fiber networks, an intriguing class of soft materials with unique non-thermal transitions associated with the onset of mechanical rigidity~\cite{Broedersz2011,Sharma2016,Bantawa2023}. Active stresses continuously actuate and anneal soft modes of the emerging network, thus perhaps driving the active solid towards a rigidity transition. Finally, the next challenge is to use optically responsive active fluids~\cite{ross2019controlling} to design programmable in-plane strains that generate folding into more complex 3D life-like forms, thus extending work on thin elastic membranes with spatially programmable swelling~\cite{Klein2007,Kim2012}. 

\section{Acknowledgements} We thank Layne Frechette, Bulbul Chakraborty, and Xiaoming Mao for insightful discussions. This work was primarily supported by the Department of Energy (DOE) DE-SC0022291 (JB, SR, IK, SF, ZD). FB acknowledges support of the Gordon and Betty Moore post-doctoral fellowship (GBMF award \#2919). The theory was supported in part by grant NSF PHY-2309135 to the Kavli Institute for Theoretical Physics (MB). V.V. and S.C. acknowledges partial support from the Army Research Office under grant W911NF-22-2-0109 and W911NF-23-1-0212, the National Science Foundation through the Center for Living Systems (grant no. 2317138), the National Institute for Theory and Mathematics in Biology, the Simons Foundation and the Chan Zuckerberg Foundation.

\section{Supplementary Information}

\renewcommand\thefigure{S\arabic{figure}}    
\setcounter{figure}{0}    

\subsection{Sample preparation}
The samples were prepared following the previously published protocol~\cite{Berezney2022}, with the main difference being the presence of fascin. Actin was purified from rabbit muscle acetone powder (Pel-Freez), incubated with the non-hydrolyzable ATP analog AMP-PNP (Roche), and flash-frozen in small aliquots ({$\approx$}\SI{100}{\micro\molar}). Lyophilized fascin (Cytoskeleton Inc.) was resuspended at \SI{92.5}{\micro\molar} and also flash-frozen before usage. Actin was fluorescently labeled using lyophilized 488-Phalloidin (Thermo Fisher). It was dissolved in DMSO at \SI{66}{\micro\molar} and stored at -20 $^\circ$C prior to usage. Actin and fascin stocks were combined and diluted to the desired concentration in an ATP-free G-buffer (2 mM HEPES, 0.1 mM CaCl2, pH 7.5 with KOH) before being added to the active fluid mix.

The active fluid mix consisted of MTs, kinesin-streptavidin (KSA) motor clusters, and PRC1, as well as ATP fuel regeneration and oxygen-scavenging systems. Most components were stored in M2B buffer (80 mM PIPES, 2 mM MGCl2, 1 mM EGTA, pH 6.8 with KOH). Tubulin from bovine brain (with 3\% of monomers labelled with Alexa Fluor 647 dye) was polymerized in the presence of GMPCPP in M2B and flash-frozen. Motor clusters were formed with biotin-K401 (expressed and purified from E. Coli) incubated with streptavidin tetramers in M2B at a molar ratio of 2:1. PRC1 was expressed and purified from E. Coli~\cite{Chandrakar2022, Subramanian2010}. The ATP system consisted of NPE-caged ATP (ThermoFisher), Pyruvate Kinase/Lactic Dehydrogenase (PK/LDH, Sigma-Aldrich), and Phospho(enol)pyruvic acid monopotassium salt (PEP, Beantown). The oxygen-scavenging system consisted of glucose, glucose oxidase, glucose catalase, and DTT. In addition to the 2 mM MgCl$_2$ present in M2B, an additional 3.3 mM MgCl$_2$ was added to the final sample mixture. 

The following components were kept at fixed final concentrations across all samples:
\begin{itemize}
\item 488-phalloidin - \SI{0.67}{\micro\molar}
\item MTs - \SI{25}{\micro\molar}
\item PRC1 - \SI{0.4}{\micro\molar}
\item Caged ATP - \SI{2.0}{\milli\molar}
\item PK/LDH - 22 units/mL (assuming 800 units/mL stock)
\item glucose - 2.82 mg/ml
\item glucose oxidase - 0.22 mg/ml
\item glucose catalase - 0.04 mg/ml
\item DTT - \SI{5.6}{\milli\molar}
\item MgCl$_2$ - \SI{5.3}{\milli\molar} (total)
\end{itemize}

Both of the Active Buffer recipes used the components and concentrations listed above, but differed in KSA and PEP concentration. Active Buffer 1 contained \SI{250}{\nano\molar} KSA and \SI{27}{\milli\molar} PEP, while Active Buffer 2 contained \SI{200}{\nano\molar} KSA and \SI{40}{\milli\molar} PEP. For the same concentrations of actin and fascin, higher PEP levels generated weaker bundling. Thus, samples prepared with Active Buffer 1 resulted in pre-bundled actin but had a shorter active lifetime, while samples prepared with Active Buffer 2 resulted in unbundled actin but had a longer active lifetime. The Active Buffer was combined with the main sample premixture to form the final suspension. An additional 10X M2B buffer was added to ensure the final suspension contained 1X M2B. The salt concentrations in the M2B/premix were sufficient to induce actin polymerization. 

The sample chamber consisted of two acrylamide-coated coverslips separated by Parafilm strips of thickness ${\sim}$\SI{0.12}{\milli\meter}~\cite{lau2009condensation}. This resulted in chambers of length \SI{18.0}{\milli\meter}, height \SI{0.12}{\milli\meter}, and variable width 1.5--\SI{8.0}{\milli\meter}. Prior to sample loading, chambers were incubated with a 3\% Pluronic solution for 10 minutes to passivate the Parafilm walls and prevent protein sticking. The chambers were subsequently rinsed with DI water and dried.   

The final mixture was loaded into sample chambers, sealed with 5-minute epoxy, and incubated for 1 hour at room temperature for actin to polymerize. Subsequently, the sample was mounted on the microscope and exposed to UV light for one minute to uncage the ATP.

\subsection{Microscopy}
Samples were imaged with with either widefield epifluorescence microscopy or spinning disk confocal (X-Light V2, CrestOptics). Widefield images were acquired with 4X air objective (NA~=~0.20). Confocal images were acquired with a sCMOS camera (Prime95B, Photometerics) and binned to obtain $804\times 804$ squared pixels.  Time series of confocal two-fluorescence-channels (647 nm for MTs, 488 nm for phalloidin-labeled actin filaments) image stacks were acquired using a 20x air objective (NA~=~0.75). Data in Fig.~\ref{fig:network-formation-slow} was acquired with a \SI{2.2}{\micro\meter} $z$-step and a \SI{5.5}{\second} time interval between stacks. Data in Fig.~\ref{fig:buckling} was acquired with a \SI{2.5}{\micro\meter} $z$-step and a \SI{11.0}{\second} time interval. 

\subsection{Analysis methods}
\textit{Rendering of 3D composite system (Fig.~\ref{fig:overview}):} The 3D image of the two fluorescent channels (488 nm for phalloidin-labeled actin, 647 nm for MTs) was rendered using Imaris software. 

\textit{In-plane velocities from widefield and confocal data (Figs.~\ref{fig:network-formation-slow},~\ref{fig:oscillations}):} To adjust for uneven illuminations over the field of view, images were divided by a background image. Confocal $z$-stacks were converted to 2D images using maximum-intensity $z$-projection. Instantaneous in-plane velocity fields were estimated using the 2D Particle Image Velocimetry (PIV) software~\cite{Thielicke2021}. For acquiring PIV on the entire sample (Fig.~\ref{fig:oscillations}A), a 4X objective was used to scan the sample with six images that were then stitched together. For the 8 mm wide samples (Fig.~\ref{fig:oscillations}D), two images were stitched together. 

\textit{Calculation of velocity correlation length (Fig.~\ref{fig:network-formation-slow}F):} We calculated the scalar autocorrelation of the velocity field $C_{vv} (\Delta x, \Delta y, t) = \langle \vec{v}(x+\Delta x,y+\Delta y,t+\Delta t)\cdot \vec{v}(x,y,t)\rangle$ using a Fast Fourier Transform. To avoid circular autocorrelation, the velocity fields were zero-padded by 1/4th of the field size on both sides of the $x$- and $y$-axes. The padded field was convolved with a copy of it with axes reversed using the SciPy function \texttt{fftconvolve} with the ``valid'' mode setting. Correlation values were locally normalized to take into account the change of correlation window size at the boundaries due to the padding. Finally, the correlation length was defined as $\ell_v = \sqrt{\int_A{C_{vv}(x,y)}dxdy}$, where $A$ is the area in the $(x,y)$ plane of lags such that $C_{vv}(x,y) \geq 0$.

\textit{Membrane surface reconstruction, thickness determination, and calculation of membrane curvatures (Fig.~\ref{fig:network-formation-slow}D and Fig.~\ref{fig:buckling}).}  To analyze the out-of-plane instantaneous conformation of the membrane, we coarse grained the porous filamentous network and extracted a continuous surface. Confocal 3D image stacks were divided by a background image to adjust uneven brightness over the field of view. Binary masks were created at each time point by hysteretic thresholding the preprocessed stacks (low and high thresholds above the background intensity level: 30\% and 50\%, respectively). Masked stacks were coarse-grained by Gaussian filtering. We convolved the thresholded actin fluorescence intensity $I(x,y,z)$ with a 3D Gaussian to obtain a smooth field $I_s(x,y,z)$. The Gaussian kernel had a width $2\sigma$ such that the number of statistical degrees of freedom of $I$ within volumes of $(2\sigma,2\sigma,L_z)$ was $30$. The local height of the membrane $h(x,y)$ was defined as the weighted average $\sum z I_s / \sum I_s $ over a volume $(\sigma,\sigma,L_z)$ centered at $(x,y)$. Then, the root-mean-square height fluctuation was calculated as $\sqrt{\langle h^2 - \langle h \rangle^2 \rangle}$ where averages are taken over the field of view. The local thickness of the membrane at $(x,y)$ was calculated as 4 standard deviations of the distribution $I(x,y,z)$ along the local normal $\hat{n} = (-\partial_x h,-\partial_y h,1)/\sqrt{1+(\nabla h)^2}$ in a volume $(2\sigma,2\sigma,L_z)$ centered at $(x,y)$. The mean and Gaussian curvatures were calculated using smooth Gaussian derivatives with kernels truncated to $4\sigma$ (first-order derivatives) and $8\sigma$ (second-order derivatives) according to the formulas

$$
 H = \frac{(1+(\partial_y h)^2)\partial_x^2 h - 2\partial_x h \partial_y h \partial_x\partial_y h +(1+(\partial_x h)^2)\partial_y^2 h}{2(1+(\nabla h)^2)^{3/2}}\,,
$$
and 
$$
K = (\partial_x^2 h \,\partial_y^2 h-(\partial_x\partial_y h)^2)/(1+(\nabla h)^2)^2\,.
$$

\textit{Network tracking and strain analysis (Fig.~\ref{fig:buckling}).}
To analyze deformation of the membrane across the entire duration of an experiment, we focused on a sample where the actin quickly formed a network structure within \SI{0.5}{h} of activation (Active Buffer~1). This allowed us to track network nodes as fiducial markers using the TrackPy package~\cite{Trackpy2025} (Video~5). For subsequent analysis we only used nodes that could be tracked for the entire duration from \SI{0.5}{h} to the end of the experiment (\SI{6}{h}).
We then triangulated these nodes in the initial time point (\SI{0.5}{h}) via a Delaunay triangulation based on the $x$-$y$ positions (Fig.~\ref{fig:strain-analysis}A). 
To measure strain, we calculated the triangulation edge lengths (in 3D space) relative to their initial length (Fig.~\ref{fig:strain-analysis}B).
Figure~\ref{fig:strain-analysis}C shows the histogram of strains for all edges and all timepoints.

\textit{Quantification of MT polarization and actin membrane velocity in $x$-$z$ plane (Fig.~\ref{fig:oscillations}E)}: 3D fluorescence images of actin and MT were acquired with a spinning disk confocal with a 40X WI objective. At each $z$-step, both the actin (488~nm) and MT (647~nm) excitation wavelengths were flashed sequentially. An entire $z$-scan took about 40 seconds. Voxel size was dx=dy=\SI{0.55}{\micro\meter}, dz=\SI{0.4}{\micro\meter}, and total image size was $442 \times 302\times 58$ \unit{\micro m}. The 3D image was cropped along the y-axis to size $442 \times 11\times 58$ \unit{\micro m} to visualize MT bundles parallel to the x-z plane. The image was then converted into a 2D image by maximum-intensity projection along the y-axis. The actin membrane velocity was computed by PIV on the 2D image sequence. The MT director field on each frame was then computed using the structure tensor method~\cite{rezakhaniha2012experimental, duclos2020topological}. The director field was converted into a polarization field by setting $p_z = \min\{n_z, -n_z\}$ above the membrane and $p_z = \max\{n_z, -n_z\}$ below the membrane. This results in all vectors $\mathbf{p}=(p_x, p_z)$ pointing into the membrane. The MT fluorescence intensities were rescaled with a linear ramp from 2 to 98 percentiles and normalized. The resulting intensity field was used as a scalar order parameter $s$. The average weighted polarization was then computed as $\langle s\mathbf{p} \rangle / \langle s \rangle$, with the average restricted to regions above and below the membrane, so that MTs inside the membrane are excluded. 

\subsection{Model and theoretical analysis}

The minimal model described in the main text was extended to 2D. Displacement $u_i(x,y,t)$ and MT-tilt direction $p_i(x,y,t)$ are 2D vector fields, where $i = x,y$ labels the vector components. The dynamics of the polarity field is  
\begin{equation} \label{eq:p-eq}
     \partial_t p_i(x,y,t) = - \frac{p_i}{\tau_\mathrm{p}} + \alpha \, \partial_t u_i.
\end{equation}
Importantly, this model only describes the large-scale MT tilting, with small-scale chaotic dynamics averaged out.
The balance of forces acting on the membrane yields an equation for the displacement dynamics
\begin{equation} \label{eq:u-eq}
    \gamma \partial_t u_i(x,y,t) = \partial_j \sigma_{\!ij} + f_0 \frac{p_i}{1 + |\mathbf{p}|/p_0}.
\end{equation}
where the elastic and (effective) viscous stresses are given by
\begin{equation}
     \sigma_{\!ij} = \sigma^\mathrm{el}_{\!ij} + \sigma^\mathrm{visc}_{\!ij} = 2 \mu u_{ij} + (B - \mu) u_{kk} \delta_{ij} + 2 \eta \partial_t u_{ij} + (\eta_\mathrm{B} - \eta) \partial_t u_{kk} \delta_{ij}.
\end{equation}
As shown below, the drag with the surrounding fluid leads to an effective scale-dependent bulk viscosity that suppresses modes with non-zero divergence.

\textit{Linear stability analysis.}
The linearized dynamics near the quiescent steady state $p_i = u_i = 0$ read
\begin{subequations}
\begin{align}
    \gamma \partial_t u_i &= \mu \nabla^2 u_i + B \partial_i \partial_j u_j + \eta \nabla^2 \partial_t u_i
    + \eta_\mathrm{B} \partial_i \partial_j \partial_t u_j + f_0 p_i,\\
    \partial_t p_i &= - \frac{p_i}{\tau_\mathrm{p}} + \alpha \, \partial_t u_i.
\end{align}
\end{subequations}
The displacement and polarity fields can be decomposed into divergence free (purely vortical) and rotation free (purely compressive/dilational) components by a Helmholtz decomposition:
\begin{subequations}
\begin{align}
    u_i = \partial_i \phi_u + \epsilon_{ij} \partial_j \psi_u, \\
    p_i = \partial_i \phi_p + \epsilon_{ij} \partial_j \psi_p, 
\end{align}
\end{subequations}
where $(\epsilon_{ij}) = \left(\begin{smallmatrix}
    0 & 1 \\ -1 & 0
\end{smallmatrix}\right)$ is the antisymmetric symbol.
Substituting this decomposition into the linearized dynamics, going to the Fourier domain $\psi \propto e^{i q_k x_k}$ and applying $i \epsilon_{ij} q_j/q^2$ yields
\begin{subequations} \label{eq:psi-eqs}
\begin{align}
    \gamma \partial_t \psi_u &= -\mu q^2 \psi_u - \eta q^2 \partial_t \psi_u + f_0 \psi_p, \\
    \partial_t \psi_p &=  -\frac{\psi_p}{\tau_\mathrm{p}} + \alpha \, \partial_t \psi_u, 
\end{align}
\end{subequations}
and an analogous pair of equations for $\phi_{u,p}$ with $\mu + B$ replacing $\mu$ and $\eta + \eta_\mathrm{B}$ replacing $\eta$. 
Notably, these equations are not coupled -- divergent and vortical modes are independent in the linearized dynamics. 

The linear equations~\eqref{eq:psi-eqs} are solved with the ansatz $\psi_{u,p} \propto e^{\lambda(q) t}$, which yields an eigenvalue problem for the growth rate $\lambda(q)$ and the dispersion relation (Fig. \ref{fig:disp-rel}). The fastest growing mode is always at $q = 0$. The marginal mode ($\mathrm{Re}\, \lambda(q_+) = 0$) is found as:
\begin{equation}
    q_+ = \frac{\sqrt{A - 1}}{\ell_\text{v-el}},
\end{equation}
where $A = \alpha f_0 \tau_\mathrm{p}/\gamma$ is the non-dimensional driving strength and $\ell_\text{v-el} = \sqrt{(\mu \tau_\mathrm{p} + \eta)/\gamma}$ is a visco-elastic length scale.
For channels narrower than $\pi/q_+$, large-scale oscillations are suppressed. The imaginary part of the growth rate $\lambda$ is always non-zero for the marginal mode
\begin{equation}
    \mathrm{Im} \, \lambda(q_+) = \tau_\mathrm{p}^{-1} \sqrt{A - 1},
\end{equation}
indicating that the membrane oscillates. Note that an oscillating mode with a finite wavenumber corresponds to a traveling wave where the wave speed is given by 
\begin{equation} \label{eq:vTW}
    v_\mathrm{TW} = \frac{\mathrm{Im} \, \lambda(q_+)}{q_+} = \frac{\sqrt{\ell_\mathrm{el}^2 + \ell_\mathrm{ev}^2}}{\tau_\mathrm{p}}
\end{equation}
We therefore expect the collective oscillations of the membrane to travel along the length of the channel as a wave. Indeed, this is observed in the simulations and experiments. Notably, the wave speed is independent of the dimensionless driving strength $A$. It is only dependent on the elasticity and viscosity of the membrane as well as the (effective) drag coefficient $\gamma$ and the MT tilt relaxation time $\tau_\mathrm{p}$.

\textit{Derivation of polar in-plane driving due to active bulk stress.} Denote the nematic order magnitude and director field in the bulk by $S$ and $\mathbf{n}$ such that the active nematic stress is $\sigma_a = \zeta S \mathbf{n} \otimes \mathbf{n}$. The traction force on the membrane is given by $\sigma_a \cdot \hat{\mathbf{z}} = (\mathbf{n} \cdot \hat{\mathbf{z}}) \mathbf{n}$, where $\hat{\mathbf{z}} = (0,0,1)$ is the unit vector in $z$-direction, normal to the membrane surface. 
Let us now parametrize $\mathbf{n} = (\cos(\phi) \sin(\theta), \sin(\phi) \sin(\theta), \cos(\theta))$ in terms of the out-of-plane angle $\theta$ and the polar (in-plane) angle $\phi$. Then the in-plane component of the traction force is $\sim \zeta S \sin(2 \theta) \, (\cos(\phi), \sin(\phi))/2$, which we can write as $\zeta \mathbf{p}_\mathrm{2D}^{}$. 
The magnitude $|\mathbf{p}_\mathrm{2D}^{}| = S \sin(2 \theta)$ of the 2D polar director field  accounts for both the tilt angle $\theta$ and the nematic order magnitude $S$.

\textit{Effect of the surrounding incompressible fluid.} Since the chamber's lateral dimensions are much larger than its height, we use the lubrication approximation. In this approximation, the pressure gradients along $z$ are neglected. We integrate the divergence across $z$ to obtain the pressure field $P$ as a function of the in-plane divergence, giving in Fourier space, to leading order in $qd$
\begin{equation}
    \tilde{P}(\mathbf{q}) = -\nu \frac{6 i q_k \tilde{v}_k (\mathbf{q}, z = 0)}{q^2 (d/2)^2},
\end{equation}
where $\nu$ is the fluid viscosity and $d$ is the chamber's height.
We assume that the fluid velocity equals the membrane's velocity in the membrane plane ($v_i(z = 0) = \partial_t u_i$) and vanishes at the walls. 
We then calculate the fluid force exerted on the membrane, given the membrane's velocity $v_i$. 
In Fourier space, the result reads
\begin{equation}
    \tilde{f}_i (\mathbf{q}) = - \frac{4\nu}{d} v_i - \frac{\nu d}{3} q^2 v_i - \frac{12 \nu}{d q^2} q_i q_j v_j,
\end{equation}
where we only kept terms to leading order in the dimensionless wavenumber $q d$.
The three terms play the role of an effective friction with friction coefficient $\gamma = 4\nu/d$, an effective shear viscosity $\eta = d\nu/3$, and an effective bulk viscosity $\eta_\mathrm{B}(q) = 12\nu/(d q^2)$. 
Notably, the ratio of bulk to shear viscosity scales as $(dq)^{-2}$. Thus, the membrane is increasingly incompressible at larger length scales, which is observed in experiments. 

Substituting the $\eta_\mathrm{B}(q)$ into the linearized equation for $\phi_u$ yields a term
\begin{equation}
    -\eta_\mathrm{B}(q) q^2 + \partial_t \phi_u = -\frac{12 \nu}{d} \partial_t \phi_u,
\end{equation}
which has the form of an effective drag force. 
The total effective drag acting on divergent modes is therefore
\begin{equation}
    \gamma_\mathrm{div} = \gamma +  \frac{12 \nu}{d} = \frac{16 \nu}{d}
\end{equation}
which is a factor 4 larger than the effective drag on vortical modes $\gamma_\mathrm{rot} = 4\nu/d$.
Put differently, the dimensionless driving strength is a factor four weaker for modes with non-zero divergence than for divergence-free ones $A_\mathrm{div}/A_\mathrm{rot} = 1/4$.
This explains why compressional/dilational deformations of the actin membrane are strongly suppressed on large length scales.

\textit{Simulations.} We simulated the PDE model using the finite-element software COMSOL Multiphysics (v6.1). To suppress the divergent part of the displacement field, we set the bulk modulus much higher than the shear modulus ($B = 100\mu$). The channel geometry was implemented as a long rectangular domain with slip-walls (allowing only tangential displacement) along the long edges and clamped boundary conditions along the short edges. For the polarization field (MT tilt), we used free (Neumann) boundary conditions.
A parameter sweep over the channel width $w$ and non-dimensional driving strength $A$ shows that the onset of self-excited large-scale waves is in excellent agreement with linear stability analysis (Fig.~\ref{fig:oscillations}H). 

In the experiments, we observe that the membrane motion is strongly suppressed near the channel ends ( Fig.~\ref{fig:oscillations}A). In simulations, chaotic waves with wavelength $\sim \lambda_\mathrm{c}$ persist along the clamped short edges of the domain. To suppress these edge waves, we ramped up the friction coefficient near the short ends of the domain to suppress the local active driving ($A < 1$). In experiments, increased effective friction near the channel ends might result from the way they were glued shut.

\newpage
\subsection{Supplementary Figures}

\begin{figure}[h]
    \centering
    \includegraphics{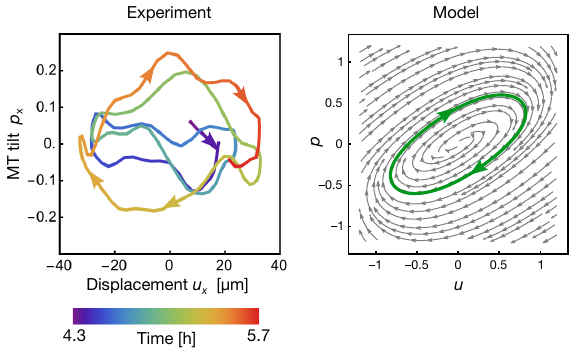}
    \caption{Limit-cycle oscillations of local displacement and MT tilt (polarization). Phase portrait of the ODE model Eq.~\eqref{eq:minimal-model} in the regime where the fixed point at $p = u = 0$ is unstable, giving rise to a limit cycle (green line). Model parameters $k_\mathrm{eff} = \alpha = p_0 = \tau_\mathrm{p} = 1$, $f_0 = 3$.}
    \label{fig:limit-cycles}
\end{figure}

\begin{figure}[h]
    \centering
    \includegraphics[scale=1.2]{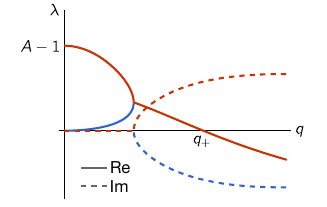}
    \caption{Characteristic dispersion relation for Eq.~\eqref{eq:psi-eqs} in the regime $A > 1$. Note that the modes near the marginal mode $q_+$ are oscillatory, as indicated by the non-zero imaginary part of the growth rate $\lambda$. These modes give rise to traveling waves; see Eq.~\eqref{eq:vTW}.
    }
    \label{fig:disp-rel}
\end{figure}

\begin{figure}
    \centering
    \includegraphics{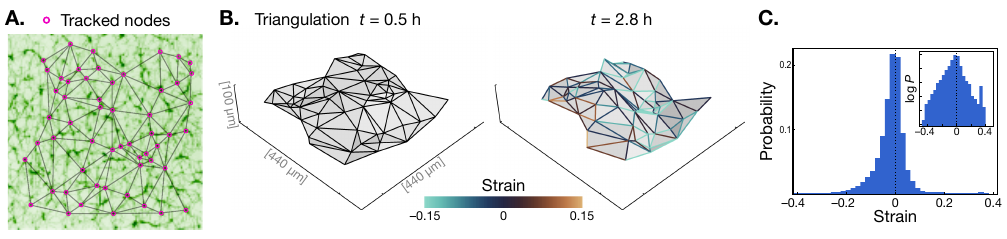}
    \caption{Actin network tracking and strain analysis.
    (A)~Tracked nodes (purple circles) and their triangulation overlaid on the actin intensity image (green shading).
    (B)~Triangulations based on the tracked nodes with node heights obtained from the midsurface extraction (cf.\ Fig.~\ref{fig:buckling}A,B). Edge color in shows the strain relative to the reference state at $t = \SI{0.5}{h}$ (left).
    (C)~Histogram of strains for all triangulation edges and all time points. Inset: Logarithmic plot shows approximately exponential distribution of strain magnitude.
    }
    \label{fig:strain-analysis}
\end{figure}

\begin{figure}
    \centering
    \includegraphics[width=0.8\textwidth]{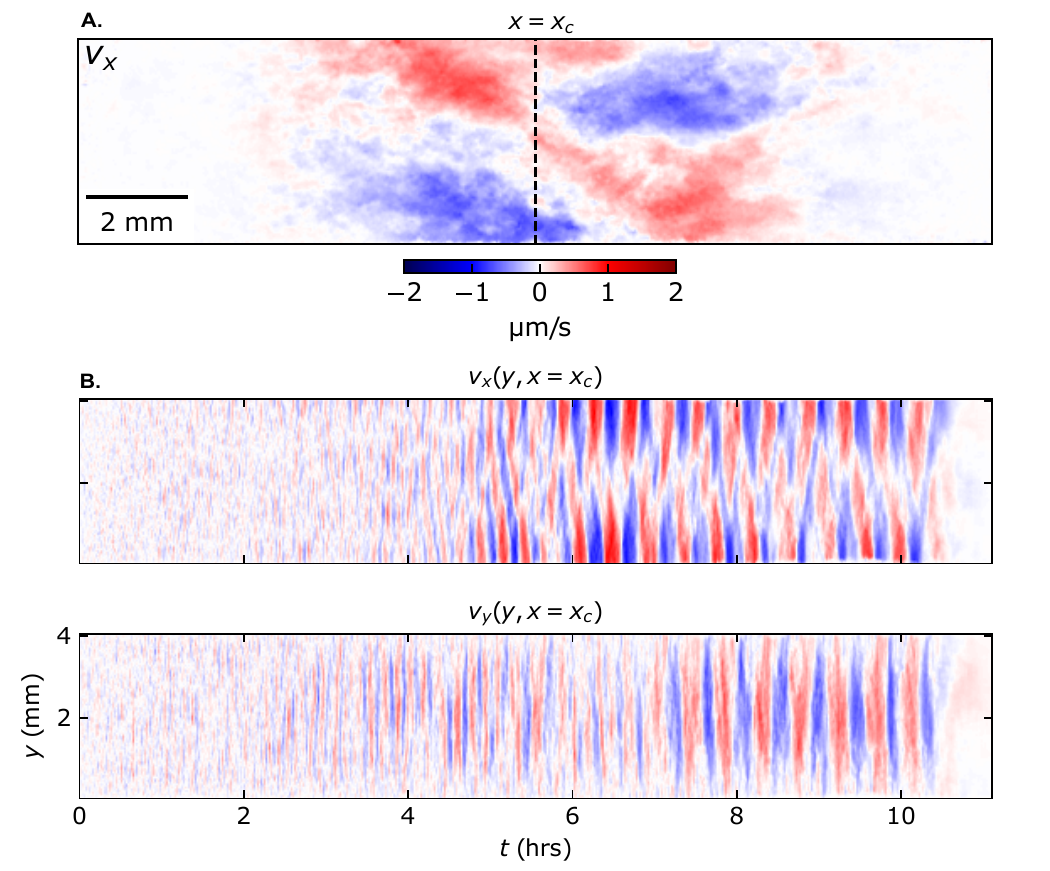}
    \caption{Gradual onset of global oscillations. (A)~Snapshot of $v_x$ component of velocity at $t{=}5.56$ hours. (B) Kymographs of the $v_x$ (top) and $v_y$ (bottom) components of velocity along the vertical line indicated in (A). Same dataset as Fig.~\ref{fig:oscillations}A--C.}
    \label{fig:oscillation-kymo}
\end{figure}

\begin{figure}
    \centering
    \includegraphics{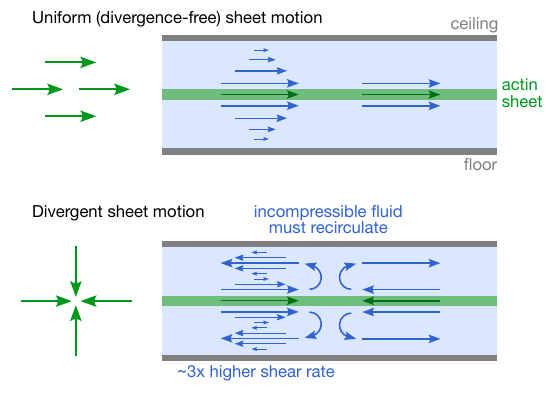}
    \caption{Illustration of effective drag forces on the membrane (green) due to the viscosity of the surrounding fluid (blue). (Top) For divergence-free motion of the membrane, the fluid is dragged along with the actin membrane, resulting in a shear rate $2v_\mathrm{mem}/d$, where $d$ is the chamber height. In contrast, divergent membrane motion forces recirculation of the incompressible surrounding fluid, which increases the shear rate by a factor of ~3, since the height across which the flow velocity drops to zero is effectively reduced by a factor 3.}
    \label{fig:effective-drag}
\end{figure}

\newpage
\subsection{Supplementary Videos}

\textbf{Video~1:}  Preformed actin-fascin bundles (green) form a network and contract into a membrane under the influence of an active MT fluid (magenta). Rendering from a 3D image acquired with a spinning disk confocal microscope. Same conditions as those in Fig.~1(A)--(E).

\textbf{Video~2:} Maximum-intensity z-projection of actin-fascin bundles forming network under influence of active MT flows (\SI{1.5}{\micro\molar} actin, \SI{0.5}{\micro\molar} fascin, Active Buffer 1). Same sample as Fig.~\ref{fig:network-formation-fast}. The average translation within the displayed region was removed to highlight network reconfiguration during periods of large in-plane motion.

\textbf{Video~3:} Actin network formation from dilute bundles (\SI{1.0}{\micro\molar} actin, \SI{0.33}{\micro\molar} fascin, Active Buffer 1). 

\textbf{Video~4:}  Active fluid induces assembly of the actin network in a sample that had initially unbundled actin (\SI{1.5}{\micro\molar} actin, \SI{1.0}{\micro\molar} fascin, Active Buffer 2). Same sample as Fig.~\ref{fig:network-formation-slow}. 

\textbf{Video~5:} Computation of in-plane strain fluctuations from tracking nodes in z-projected images of the actin-fascin network (\SI{1.5}{\micro\molar} actin, \SI{0.5}{\micro\molar} fascin, Active Buffer 1). Same sample as Fig.~\ref{fig:buckling}. 

\textbf{Video~6:}  Emergence of system-sized shear oscillations in the actin network (\SI{3.0}{\micro\molar} actin, \SI{2.2}{\micro\molar} fascin, Active Buffer 2). Same sample as Fig.~\ref{fig:oscillations}A-C.  

\textbf{Video~7:}  Width dependence of system-sized shear oscillations (\SI{6.0}{\micro\molar} actin, \SI{2.0}{\micro\molar} fascin, Active Buffer 1). Same sample as Fig.~\ref{fig:oscillations}E-F. 

\textbf{Video~8:} Velocity field (arrows) and vorticity (color) showing self-excited waves in a simulation with $w/\ell_\text{v-el} = 15, A = 1.6$ and channel length $L = 4 w$.

\setcounter{figure}{0}
\renewcommand{\thefigure}{S\arabic{figure}}

\bibliography{main}

\end{document}